\documentstyle[preprint,pra,aps]{revtex}
\tightenlines
\begin{document}
\draft

\title{Lens optics as an optical computer for group contractions}

\author{S. Ba{\c s}kal \footnote{electronic address:
baskal@newton.physics.metu.edu.tr}}

\address{Department of Physics, Middle East Technical University,
06531 Ankara, Turkey}

\author {Y. S. Kim \footnote{electronic address:
yskim@physics.umd.edu}}
\address{ Department of Physics, University of Maryland, College Park,
Maryland 20742}

\maketitle

\begin{abstract}
It is shown that the one-lens system in para-axial optics can serve as
an optical computer for contraction of Wigner's little groups and
an analogue computer which transforms analytically computations on a
spherical surface to those on a hyperbolic surface.  It is shown
possible to construct a set of Lorentz transformations which leads
to a two-by-two matrix whose expression is the same as those in
the para-axial lens optics.  It is shown that the lens focal condition
corresponds to the contraction of the $O(3)$-like little group for a
massive particle to the $E(2)$-like little group for a massless
particle, and also to the contraction of the $O(2,1)$-like little
group for a space-like particle to the same $E(2)$-like little group.
The lens-focusing transformations presented in this paper allow us to
continue analytically the spherical $O(3)$ world to the hyperbolic
$O(2,1)$ world, and vice versa.

\end{abstract}


\section{Introduction}\label{intro}
The six-parameter Lorentz group was initially introduced to physics as
a group of Lorentz transformations applicable to the four-dimensional
Minkowskian space.  However, the Lorentz group can serve as the basic
mathematical language for many branches of physics.  It serves as the
backbone for the theory of coherent and squeezed states of
light~\cite{yuen76,knp91}.
Recently we are realizing that the Lorentz group can serve as the
standard language for
classical ray optics, including polarization optics~\cite{hkn97},
interferometers~\cite{hkn00},
layer optics~\cite{monzon01,geo01}, lens optics~\cite{baskal01},
and cavity optics~\cite{baskal02}.

Since the Lorentz group provides the underlying scientific
language to classical optics, it is not
unreasonable to examine whether we can construct optical devices which
will perform computations in the Lorentz group.  This group has many
subgroups, and we are particularly interested in Wigner's little
groups which dictate the internal space-time symmetries of
relativistic
particles~\cite{wig39}.  While these groups play the fundamental role
in
particle physics, they had a stormy history in connection with their
role in explaining the space-time symmetry of massless
particles~\cite{kiwi90jm}.

Wigner's little group is defined to be the maximal subgroup of the
Lorentz group whose transformations leave the four-momentum of a given
particle invariant.
The little groups for massive, massless, and space-like momentum are
like $O(3)$, $E(2)$, and $O(2,1)$ respectively~\cite{like}.   The
$O(3)$ group is the
three-dimensional rotation group and can provide computations on the
numbers distributed on a spherical surface. The $O(2,1)$ group
provides
transformations in the Minkowskian space of two space-like and one
time-like dimensions.  Thus, this group deals with the numbers on
a hyperbolic surface.  The $E(2)$ group stands for Euclidean
transformations
on a flat surface.  It consists of two translational degrees of
freedom
as well as the rotation around the origin.

The transitions from
$O(3)$ to $E(2)$, and from $O(2,1)$ to $E(2)$ are called the group
contractions in the literature~\cite{inonu53,knp86}, and they are
known
to be singular transformations which forbid analytic continuation from
$O(3)$
to $E(2)$.  After the $O(3)$ or $O(2,1)$ is contracted to $E(2)$, it
is not possible to recover either of the two groups from $E(2)$.
In addition, the little groups are not exactly the $O(3), E(2)$, and
$O(2,1)$ groups whose geometry is quite transparent to us.  They are
only ``like''\cite{like}.  The question is then whether the contraction
of $O(3)$ to $E(2)$ necessarily mean the contraction of the $O(3)$-like
little group to $E(2)$-like little group.  This conceptual question
also has been discussed extensively in the
literature~\cite{kiwi90jm,knp86,bacry68}.
In this paper, we only use the results which can be represented in
the two-by-two matrix representation of the Lorentz group.

The one-lens system consists of one lens matrix and two translation
matrices~\cite{yariv75}.  The combined matrix can be written in terms
of the two-by-two matrices corresponding to four-by-four
Lorentz-transformation matrices which constitute the transformations
of the little groups~\cite{baskal01}.  However,
unlike the case of the little groups, the parameters of the
two-by-two matrices are analytic, especially in the neighborhood of
the focal condition in which the upper-right element vanishes.  On
the other hand, from the little group point of view, this is
precisely where the group contraction occurs, and this transformation
is singular as was mentioned above.  Then how can we establish the
correspondence between singular and non-singular representations?

Indeed, if we can represent those three little groups using one convex
lens, the result would be quite interesting, especially
in view of the fact that computations in a hyperbolic world can be
performed in a spherical world, and vice versa.
In order to achieve this goal, we have to develop a mathematical
device which establishes a bridge between lens optics and the little
groups, starting from Wigner's original idea of finding the maximal
subgroups of the Lorentz group which will leave the
four-momentum of a given particle invariant.

With this point in mind, we present a different set of Lorentz
transformations
achieving the same purpose~\cite{hk88}.  We then show that the
representation of the little groups in this set coincides with the matrix
representation of the one-lens system.
It is then seen that the focal condition corresponds to the transition
from one little group to another.  The transition is analytic.  In
this way, we achieve an analytic transformation of computations on a
hyperbolic surface to a spherical surface.

In this paper, we are employing many sophisticated mathematical
items such as the Lie algebra, compact groups, non-compact groups,
solvable groups, as well as group contractions.  However, we are very
fortunate to be able to avoid these words and get directly into the
computational world using only familiar two-by-two matrices
without complex numbers.  Our mathematics starts from the well-known
two-by-two matrix formulation of the one-lens system.

In Sec.~\ref{formul}, we start with one lens matrix and two
translation
matrices, and derive a core matrix to be studied in detail.
In Sec.~\ref{little}, we introduce Wigner's little groups and their
traditional two-by-two representations, and point out that they are 
not suitable
for describing the core matrix in the one-lens system because the
transition from one little group to another is a singular
transformation.
In Sec.~\ref{contrac}, for the little groups, we introduce a different
set of
Lorentz transformations which can serve as a bridge between the
symmetries of relativistic particles and the one-lens system.
It is noted that the transition from one little group to another can
be achieved analytically.  It is noted that the group contraction is
not always a singular transformation.

Then in Sec.~\ref{lenscon}, we formulate the one-lens system in terms
of the little groups and the analytic group contraction.
In Sec~\ref{cavity}, it is pointed out that the cavity optics is a
special
case of the one-lens system, but that this simplified system contains
all the essential
features of group contractions.  In Sec.~\ref{comp}, it is shown how the 
abstract idea of group contractions leads to a tool of concrete numerical
calculations which can be carried out by optical devices.

\section{Formulation of the Problem}\label{formul}

The simplest lens system is of course the one-lens system
with the lens matrix
\begin{equation}
\pmatrix{1 & 0 \cr -1/f & 1} ,
\end{equation}
where $f$ is the focal length.  We assume that the focal length is
positive throughout the paper.  The translation matrix takes
the form
\begin{equation}
\pmatrix{1 & d \cr 0 & 1} .
\end{equation}
If the object and image are $d_{1}$ and $d_{2}$ from the lens
respectively, the optical system is described by
\begin{equation}\label{lens11}
\pmatrix{1  & d_{2} \cr 0 & 1} \pmatrix{1  & 0 \cr -1/f  & 1 }
\pmatrix{1  & d_{1} \cr 0 & 1} .
\end{equation}
The multiplication of these matrices leads to
\begin{equation}
\pmatrix{1 - d_{2}/f  & d_{1} + d_{2} - d_{1} d_{2}/f
      \cr - 1/f & 1 - d_{1}/f} .
\end{equation}
The image becomes focused
when the upper right element of this matrix vanishes with
\begin{equation}\label{focal}
{ 1 \over d_{1}} + { 1 \over d_{2}} = { 1 \over f}.
\end{equation}

The problem with this expression is that the off-diagonal elements are
not dimensionless, but it can be decomposed into
\begin{equation}\label{off11}
\pmatrix{\left(d_{1}d_{2}\right)^{1/4}  & 0 \cr 0  &
 \left(d_{1}d_{2}\right)^{-1/4}  }
\pmatrix{1 - x_{2}  &  2\cosh\rho - x \cr
 - x & 1 - x_{1}}
\pmatrix{\left(d_{1}d_{2}\right)^{-1/4} & 0
\cr 0 & \left(d_{1}d_{2}\right)^{1/4} } ,
\end{equation}
with
\begin{eqnarray}
&{}&   x_{1} =  {d_{1} \over f },  \qquad
       x_{2} = { d_{2} \over f } , \qquad
       x =  {\sqrt{d_{1} d_{2}} \over f },  \nonumber \\[2ex]
&{}& \cosh\rho = {1 \over 2} \left(\sqrt{d_{1}/d_{2}} +
\sqrt{d_{2}/d_{1}}\right).
\end{eqnarray}
The matrix in the middle, the core matrix, can now be written as
\begin{equation}\label{core-}
\pmatrix{1 - x_{2}  & 2 \cosh\rho - x \cr - x & 1 - x_{1}} .
\end{equation}
In the camera configuration, both the image and object distances are
larger than the focal length, and both $(1 - x_{1})$ and $(1 - x_{2})$
are negative.  Thus we start with the negative of the above matrix
\begin{equation}\label{core}
\pmatrix{x_{2} - 1  & x - 2 \cosh\rho \cr x & x_{1} - 1} .
\end{equation}

We can further renormalize this matrix to make the two diagonal
elements
equal.  For this purpose, we can write it as
\begin{equation}
\pmatrix{b & 0 \cr 0 & 1/b}
\pmatrix{z - 1 &  x - 2\cosh\rho  \cr x & z - 1}
\pmatrix{b & 0 \cr 0 & 1/b} ,
\end{equation}
with $$
b = \left( {x_{2} - 1 \over x_{1} - 1 } \right)^{1/4} .
$$
Then the core matrix becomes
\begin{equation}\label{core+}
\pmatrix{z - 1  & x - 2 \cosh\rho \cr x & z - 1} ,
\end{equation}
with
\begin{equation}
z = 1 + \sqrt{(x_{1} - 1)(x_{2} - 1)} .
\end{equation}
In terms of the $\rho$ and $x$ variables, $z$ can be written as
\begin{equation}
z = 1 + \sqrt{x^{2} - 2x\cosh\rho + 1} .
\end{equation}

We shall use the core matrix of Eq.(\ref{core+}) as the starting point
in this paper.
If $x$ is smaller than $2(\cosh\rho)$, the core matrix can be written
as
\begin{equation}\label{lgo3}
\pmatrix{\cos(\phi/2) &  - e^{-\eta/2} \sin(\phi/2)
\cr e^{\eta/2} \sin(\phi/2) & \cos(\phi/2)} ,
\end{equation}
where the range of the angle variable $\phi$ is between $0$ and $\pi$,
and $\eta$ is positive.

This form of the core matrix serves a very useful purpose in
laser optics which consists of chains of the one-lens
system~\cite{yariv75,haus84,hawk95}.  Its connection with the Lorentz
group and Wigner rotations has been studied recently by the
present authors~\cite{baskal02}.  Indeed, this is the starting point
of this paper, where we intend to establish a connection between
the one-lens system and a set of Lorentz transformations.

If $x = 2\cosh\rho$, the above expression becomes
\begin{equation}\label{lge2}
\pmatrix{1 & 0 \cr 2\cosh\rho & 1},
\end{equation}
and the focal condition of Eq.(\ref{focal}) is satisfied.
If $x$ is is greater than $2(\cosh\rho)$, all the elements in the core
matrix of Eq.(\ref{core+}) become positive.  Thus, it is appropriate
to write it as
\begin{equation}\label{lgo21}
 \pmatrix{\cosh(\chi/2) & e^{-\eta/2} \sinh(\chi/2) \cr
 e^{\eta/2} \sinh(\chi/2) & \cosh(\chi/2) } .
\end{equation}

As we shall see in Sec.~\ref{little}, the expressions given in
Eq.(\ref{lgo3}), Eq.(\ref{lge2}), and Eq.(\ref{lgo21}) take the same
mathematical forms as those of the representations of the $O(3),
E(2)$, and $O(2,1)$-like little groups.  The transition from one to 
another form is a singular transformation. On the other hand, the core 
matrix of Eq.(\ref{core+}) is analytic in the $x$ and $\rho$ variables 
when both $x_{1}$ and $x_{2}$ are greater than 1.  We are thus led to 
look for another set of Lorentz transformations with analytic parameters.
This will enable us to write those transformation parameters in terms
of the lens parameters of Eq.(\ref{core+}).  In so doing, we can
establish a correspondence between lens optics and the transformations
of the little groups, and we can achieve transformations from one
little group to another by adjusting focal conditions.

\section{Little Groups}\label{little}

In his 1939 paper on the Lorentz group~\cite{wig39}, Wigner considered
the maximum subgroup of the Lorentz group whose transformations leave
the four-momentum of a given free particle invariant. This subgroup is
called Wigner's little group.  Wigner observed that there are three
classes
of the little group.  In the Minkowskian space of the space-time
coordinate $(t, z, x, y)$, the four-vector
\begin{equation}\label{mom22}
m(1, 0, 0, 0)
\end{equation}
corresponds to the four-momentum of a massive particle at rest.  To
this four-vector, we can apply
three-dimensional rotation matrix, like the rotation matrix around
the $y$ axis:
\begin{equation}\label{m441}
\pmatrix{1 & 0 & 0 & 0 \cr 0 & \cos\phi & -\sin\phi & 0 \cr
   0 & \sin\phi & \cos\phi & 0 \cr 0 & 0 & 0 & 1} ,
\end{equation}
without changing the four-momentum of Eq.(\ref{mom22}).
In optics, it is more convenient to use the two-by-two representation
this matrix~\cite{hkn00}.  The rotation matrix then becomes
\begin{equation}\label{m200}
\pmatrix{\cos(\phi/2) & -\sin(\phi/2) \cr \sin(\phi/2) & \cos(\phi/2)
} .
\end{equation}

In order to study the little group for a particle moving along the
$z$ direction, we can start with a particle with
four-momentum~\cite{hks86jm}
\begin{equation}\label{mom11}
  m (\cosh\eta, -\sinh\eta, 0, 0) .
\end{equation}
This particle moves in the negative $z$ direction with the speed of
$c (\tanh\eta)$.  To this four-vector, if we apply the boost matrix
\begin{equation}\label{m442}
\pmatrix{\cosh\eta & \sinh\eta & 0 & 0 \cr
 \sinh\eta & \cosh\eta & 0 & 0 \cr 0 & 0 & 1 & 0 \cr 0 & 0 & 0 & 1} ,
\end{equation}
the four-vector returns to the form given in
Eq.(\ref{mom22}).  Here again, it is more convenient to
use the two-by-two representation of this boost matrix which takes
the form~\cite{hks86jm}
\begin{equation}\label{m201}
\pmatrix{e^{\eta/2}  & 0 \cr 0 & e^{-\eta/2}} .
\end{equation}

Thus, in order to construct a representation of the little group for
the four-momentum of Eq.(\ref{mom11}), we boost it to that of
Eq.(\ref{mom22}) using the boost matrix of Eq.(\ref{m442}) or
(\ref{m201}), perform the rotation of Eq.(\ref{m441}) or (\ref{m200})
which
does not change the momentum, and then boost the momentum back to
the original form of Eq.(\ref{mom11}).  In the two-by-two
representation,
this chain of matrices take the form
\begin{equation}\label{m202}
\pmatrix{e^{-\eta/2}  & 0 \cr 0 & e^{\eta/2}}
\pmatrix{\cos(\phi/2) & -\sin(\phi/2) \cr
 \sin(\phi/2) & \cos(\phi/2) }
\pmatrix{e^{\eta/2}  & 0 \cr 0 & e^{-\eta/2}}  .
\end{equation}
After the multiplication, the result becomes
\begin{equation}\label{m203}
\pmatrix{\cos(\phi/2) & - e^{-\eta} \sin(\phi/2) \cr
  e^{\eta} \sin(\phi/2) & \cos(\phi/2)} .
\end{equation}
The mathematical form of this matrix is identical to that of 
Eq.(\ref{lgo3}).

This is the reason why the little groups can play a role in lens
optics and vice versa.  The group represented in this way is called
the $O(3)$-like little group for a massive particle~\cite{hks86jm}.
As was mentioned in Sec.~\ref{formul}, this expression is also one
of the starting formulas in laser
optics~\cite{baskal02,yariv75,haus84,hawk95}.

What happens if the four-momentum is light-like?  The four-momentum
in this case is
\begin{equation}\label{mom00}
\omega(1, 1, 0, 0) .
\end{equation}
The light-like particles cannot be brought to its rest frame, and thus
cannot be brought to the form of Eq.(\ref{mom22}).  It is clear that
this four-vector is invariant under rotations around the $z$ axis.
In addition, Wigner  observed in his original paper that there are
two additional transformations which leave this light-like
four-momentum
invariant.  These matrices are extensively discussed in the
literature,
and the result is that they correspond to the form
\begin{equation}\label{iwa11}
\pmatrix{1 & 0 \cr u & 1} ,
\end{equation}
where $u$ is a complex parameter with two real independent parameters.
Since, we will be dealing with real matrices in this paper, $u$
represents only one real number.  It is interesting to note that
this form is identical to that of Eq.(\ref{lge2}).   This aspect of
the little group also has been discussed in the
literature~\cite{hks86jm}.

The theory of the little group includes also the form
\begin{equation}\label{iwa22}
\pmatrix{1 & u \cr 0 & 1} ,
\end{equation}
but it does not play a role in this paper.  This form may be useful
if we consider the case when the lower left element of the core
matrix of Eq(\ref{core+}) vanishes.  The group represented either
in the form of Eq.(\ref{iwa11}) or Eq.(\ref{iwa22}) is called the
$E(2)$-like little group for massless particles.

The little-group matrix of Eq.(\ref{iwa11}) is invariant under the
Lorentz boost along the $z$ direction, as can be seen from
\begin{equation}
\pmatrix{e^{-\eta/2}  & 0 \cr 0 & e^{\eta/2}}
\pmatrix{1  & 0 \cr u & 1 }
\pmatrix{e^{\eta/2}  & 0 \cr 0 & e^{-\eta/2}}
 = \pmatrix{1  & 0 \cr u & 1 } .
\end{equation}

There are no particles in nature with space-like four-momentum, whose
four-vector may be written as~\cite{kiyeh92jm}
\begin{equation}\label{mom55}
m(0, 1, 0, 0) ,
\end{equation}
but it occupies an important position in group theory~\cite{wig39}.
It
will become more important as it finds its place in optical sciences.
This four-vector is also invariant under rotations around the $z$
axis.
In addition, it remains invariant under boosts along the $x$ and $y$
directions.  The boost matrix along the $x$ direction takes the form
\begin{equation}\label{m208}
\pmatrix{\cosh(\chi/2) & \sinh(\chi/2) \cr
\sinh(\chi/2) & \cosh(\chi/2)} .
\end{equation}
If we apply the same Lorentz boosts as we did in two previous little
groups, the little group matrix should become
\begin{equation}\label{m211}
\pmatrix{\cosh(\chi/2) &   e^{-\eta}\sinh(\chi/2) \cr
e^{\eta}\sinh(\chi/2) & \cosh(\chi/2)} .
\end{equation}
This form is identical to Eq.(\ref{lgo21}).

We have thus far shown that the transformation matrices of the
little groups and the one-lens system take the same form.  However,
there is one crucial problem.  In the case of the core matrix
of Eq(\ref{core}), the sign change of the upper-right element can be
done analytically, but this is not true for the little group
representation.   The transition to Eq.(\ref{iwa11}) either from
Eq.(\ref{m203}) and from Eq.(\ref{m211}) is possible and is known as
the group contraction in the literature.  However, in both cases,
the two independent parameters collapse into one independent
parameter.  Thus, the inverse transformation is not possible.  This
keeps us from continuing analytically from Eq.(\ref{m203}) to
Eq.(\ref{m208}).  What should we do?

\section{Contractions of the Little Groups}\label{contrac}

In order to circumvent the singularity problem mentioned in the
preceding section, we are interested in finding a set of Lorentz
transformations which will remain analytic as we go through the
transition point where the upper-right element vanishes.  Let us
restate the problem.

If $x$ is smaller than $2\cosh\rho$, the upper-right element of the
core
matrix of Eq.(\ref{core+}) is negative while the remaining three are
positive, and it can be written in the form of Eq.(\ref{lgo3}).
If it is greater than $2\cosh\rho$, all the elements are positive,
and the core matrix should be written as Eq.(\ref{lgo21}).  There is
a value zero between these two values, which corresponds to the
focal condition.  This is precisely the point where the expressions
Eq.(\ref{lgo3}) and Eq.(\ref{lgo21}) become singular.  The purpose of
this section is to establish the connection between the little groups
and the one-lens system without this singularity.  In the computer
language, this singularity means a memory loss.

We are thus interested in a different set of Lorentz transformations
for the little groups.  We note here again that the little group
consists of transformations which leave the four momentum of a given
particle invariant~\cite{wig39}.  In order to find the set of
transformations which will bring back the four-momentum of
Eq.(\ref{mom11}) to itself~\cite{hk88}, let us first rotate the
four-momentum by
$\theta$, using the rotation matrix
\begin{equation}\label{rot22}
\pmatrix{\cos(\theta/2) & -\sin(\theta/2)
\cr \sin(\theta/2) & \cos(\theta/2)}
\end{equation}
to Eq.(\ref{mom11}). Then the four-momentum becomes
\begin{equation}\label{mom33}
  m (\cosh\eta,  (\sinh\eta)\cos\theta ,  -(\sinh\eta)\cos\theta, 0) .
\end{equation}
This four-momentum can be boosted along the $x$ direction, which then
becomes
\begin{equation}\label{mom44}
  m (\cosh\eta,  (\sinh\eta)\cos\theta , (\sinh\eta)\cos\theta, 0) ,
\end{equation}
with the boost matrix
\begin{equation}
\pmatrix{\cosh\lambda & \sinh\lambda \cr \sinh\lambda & \cosh\lambda}
.
\end{equation}
We can return to the four-momentum of Eq.(\ref{mom11}), by applying
again
the rotation matrix of Eq.(\ref{rot22}).  The net effect is
\begin{equation}
\pmatrix{\cos(\theta/2) & -\sin(\theta/2) \cr \sin(\theta/2) &
\cos(\theta/2)}
\pmatrix{\cosh\lambda & \sinh\lambda \cr \sinh\lambda & \cosh\lambda}
\pmatrix{\cos(\theta/2) & -\sin(\theta/2) \cr \sin(\theta/2) &
\cos(\theta/2)} ,
\end{equation}
which becomes
\begin{equation}\label{m204}
\pmatrix{\cosh\lambda \cos\theta
    & -\cosh\lambda \sin\theta + \sinh\lambda \cr
  \cosh\lambda \sin\theta + \sinh\lambda  & \cosh\lambda \cos\theta }
.
\end{equation}

Indeed, these two different ways of returning to the same
four-momentum
should give the same effect.  Thus, the effect of Eq.(\ref{m203}) and
and that of Eq.(\ref{m204}) are the same, and
\begin{equation}\label{m209}
\pmatrix{\cos(\phi/2) & -e^{-\eta} \sin(\phi/2) \cr
  e^{\eta} \sin(\phi/2) & \cos(\phi/2)}
= \pmatrix{\cosh\lambda \cos\theta
    & - \cosh\lambda \sin\theta + \sinh\lambda \cr
 \cosh\lambda \sin\theta + \sinh\lambda  & \cosh\lambda \cos\theta } ,
\end{equation}
with
\begin{eqnarray}\label{eq01}
&{}& \cos(\phi/2) = \cosh\lambda \cos\theta, \nonumber \\[2ex]
&{}& e^{-2\eta} = {\cosh\lambda \sin\theta - \sinh\lambda  \over
         \cosh\lambda \sin\theta + \sinh\lambda} .
\end{eqnarray}
Conversely, $\lambda$ and $\theta$ can be written in terms of
$\phi$ and $\eta$ as
\begin{eqnarray}
&{}& \cosh\lambda =
 (\cosh\eta)\sqrt{1 - \cos^{2}(\phi/2)\tanh^{2}\eta} , \nonumber
\\[2ex]
&{}& \cos\theta = { \cos(\phi/2) \over (\cosh\eta)
      \sqrt{1 - \cos^{2}(\phi/2)\tanh^{2}\eta} } .
\end{eqnarray}
This leads to
\begin{equation}
\cosh\lambda = { \cosh\eta \over
                 \sqrt{1 + (\sinh^{2}\eta) \cos^{2}\theta} },
\end{equation}
which means that the boost parameter $\lambda$ is determined from
the rotation angle $\theta$ for a given value of the boost parameter
$\eta$.

The above relations are valid only when $(\cosh\lambda \sin\theta)$ is
greater than $\sinh\lambda$.
Otherwise, instead of Eq.(\ref{m202}), we have to start from
\begin{equation}\label{m206}
\pmatrix{e^{-\eta/2}  & 0 \cr 0 & e^{\eta/2}}
\pmatrix{\cosh(\chi/2) & \sinh(\chi/2) \cr
 \sinh(\chi/2) & \cosh(\chi/2) }
\pmatrix{e^{\eta/2}  & 0 \cr 0 & e^{-\eta/2}},
\end{equation}
which leads to
\begin{equation}\label{m210}
 \pmatrix{\cosh(\chi/2) & e^{-\eta} \sinh(\chi/2) \cr
 e^{\eta} \sinh(\chi/2) & \cosh(\chi/2) } .
\end{equation}
This form is identical to Eq.(\ref{lgo21}), and should also be equal 
to Eq.(\ref{m204}).  We write this as
\begin{equation}
 \pmatrix{\cosh(\chi/2) & e^{-\eta} \sinh(\chi/2) \cr
 e^{\eta} \sinh(\chi/2) & \cosh(\chi/2) }  =
 \pmatrix{\cosh\lambda \cos\theta
    & -\cosh\lambda \sin\theta + \sinh\lambda \cr
  \cosh\lambda \sin\theta + \sinh\lambda  & \cosh\lambda \cos\theta }
,
\end{equation}
which leads to the identities
\begin{eqnarray}\label{eq02}
&{}& \cosh(\chi/2) = \cosh\lambda \cos\theta ,  \nonumber \\[2ex]
&{}&  e^{-2\eta} = -\left( {\cosh\lambda \sin\theta - \sinh\lambda
\over
  \cosh\lambda \sin\theta +\sinh\lambda} \right) .
\end{eqnarray}
Conversely,
\begin{eqnarray}
&{}& \cosh\lambda =
(\cosh\eta)\sqrt{\cosh^{2}(\chi/2) - \tanh^{2}\eta} \, , \nonumber
\\[2ex]
&{}& \cos\theta = {\cosh(\chi/2) \over
 (\cosh\eta)\sqrt{\cosh^{2}(\chi/2) - \tanh^{2}\eta} }  .
\end{eqnarray}
In this case, the boost parameter $\lambda$ takes the form
\begin{equation}
\cosh\lambda = {\sinh\eta \over \sqrt{\cosh^{2}\eta \cos^{2}\theta -
1}} .
\end{equation}
Here, the boost parameter $\lambda$ is determined by the little group
parameter $\theta$ for a given value of $\eta$.

While the quantity
\begin{equation}\label{12elem}
 {\cosh\lambda \sin\theta - \sinh\lambda  \over
  \cosh\lambda \sin\theta +\sinh\lambda}
\end{equation}
changes the sign from $(plus)$ to $(minus)$, it has to go through
zero.
With the parameters $\lambda$ and $\theta$, this process is quite
analytic.  On the other hand, the exponential factor
$\exp(-2\eta)$
is always positive.  Thus, changing $\exp(-2\eta)$ to $-\exp(-2\eta)$
cannot be achieved analytically.  This is necessarily a singular
transformation.  However, this exponential factor becomes vanishingly
small when $\eta$ becomes very large.  Perhaps we are allowed to
change
the sign when it is vanishingly small, but this is still a
non-analytic
continuation.  Furthermore, let us look at the expressions given in
Eq.(\ref{m203}) and Eq.(\ref{m210}).  This sign change is accompanied
by the transition of a rotation matrix
of the form of Eq.(\ref{m200})
to a boost matrix of the form given in Eq.(\ref{m208}).

Indeed, by changing the parameters from $\phi$ and $\eta$ to $\theta$
and $\lambda$, we can analytically navigate through the vanishing
value
of the upper-right element of matrices of Eq.(\ref{m209}).  The
process
of approaching this zero value either from the positive or negative
side is called the group contraction in the literature.  In this
paper,
however, we are eventually interested in how these parameters operate
in
lens optics.  We shall come back to this issue in Sec.~\ref{lenscon}.

\section{Lens Optics and Group Contractions}\label{lenscon}
In Sec.~\ref{formul}, we started with a camera-like one-lens system,
and
derived
\begin{equation}
\pmatrix{z - 1  & x - 2\cosh\rho \cr  x  &  z -1} =
\pmatrix{ \cos(\phi/2) & -e^{-\eta}\sin(\phi/2)
      \cr e^{\eta} \sin(\phi/2)  & \cos(\phi/2)} ,
\end{equation}
for $x$ smaller than $2(\cosh\rho)$, and $x$ is positive.
Here all the parameters are determined from $ d_{1}, d_{2}$, and $f$
of
the lens optics. If we gradually increase the value of $x$, the
upper-right element becomes zero and then positive.  The right-hand
side
of the above expression cannot accommodate this transition.

The right-hand side is a familiar expression both in optics and
the Lorentz group.  In Sec.~\ref{little}, we started with Wigner's
little
groups, and noted that the expressions given in Eq.(\ref{lgo3}) and
Eq.(\ref{m203}) are identical to each other.  The parameters in
Eq.(\ref{m203}) are the Lorentz-transformation parameters for
Wigner's $O(3)$-like little group for massive particles.

In order to circumvent the above-mentioned singularity problem, we
have
chosen a different set of Lorentz-transformation parameters, and the
result was the expression given in Eq.(\ref{m209}).  In terms of these
parameters, the core matrix can be written as
\begin{equation}\label{m205}
\pmatrix{z - 1  & x - 2\cosh\rho \cr x & z - 1} =
\pmatrix{\cosh\lambda \cos\theta
    & -\cosh\lambda \sin\theta + \sinh\lambda \cr
  \cosh\lambda \sin\theta + \sinh\lambda  & \cosh\lambda \cos\theta }
.
\end{equation}
Here, both sides have the upper-right their upper-right elements
which are analytic as they go through zero.

The parameters are now related by
\begin{eqnarray}
&{}&   x - 2\cosh\rho = \sinh\lambda -
            \cosh\lambda \sin\theta , \nonumber \\[2ex]
&{}&   x = \sinh\lambda + \cosh\lambda \sin\theta ,
\end{eqnarray}
and therefore to
\begin{eqnarray}
&{}& \sinh\lambda = x - \cosh\rho , \nonumber \\[2ex]
&{}& \sin\theta = {\cosh\rho \over \sqrt{1 + (x - \cosh\rho)^{2}}} .
\end{eqnarray}
We are thus able to write the Lorentz-transformation parameters
$\lambda$ and $\theta$ in terms of the parameters of the one-lens
system.

Thus, by adjusting the lens parameters, we can now perform
transformations in Wigner's little groups.  It is interesting to note
that we perform group contractions whenever we try to focus the object
before taking a camera photo.  Unlike the traditional procedures,
the contraction presented this paper is an analytic transformation,
which provides a reversible process from Eq.(\ref{lgo3}) to
Eq.(\ref{lgo21}) through Eq.(\ref{lge2}).  What significance does
this have?  We shall return to this question in Sec.~\ref{comp}.

\section{Cavity Optics}\label{cavity}
In our previous paper~\cite{baskal02}, we studied light beams in laser
cavities.  One cavity cycle there consists of two lenses with the same
image and object distances.  We are thus led to consider the one-lens
system with $d_{1} = d_{2} = d $, and thus
\begin{equation}
x_{1} = x_{2} = x .
\end{equation}

The core matrix of Eq.(\ref{core+}) becomes

\begin{eqnarray}
&{}&  x - 2 = \sinh \lambda -\cosh \lambda \sin \theta ,   \nonumber
\\
&{}& x= \sinh \lambda + \cosh \lambda \sin \theta .
\end{eqnarray}
Therefore, $ \cosh \lambda \sin \theta = 1 $,  or
\begin{equation}
\sin \theta = {1 \over \cosh \lambda },
\end{equation}
which is satisfied by the physical values of $\theta$ and $\lambda$.
Furthermore, this relation reduces Eq.(\ref{m205}) to
\begin{equation}\label{m207}
\pmatrix{x - 1  & x - 2 \cr x & x - 1}
= \pmatrix{ \sinh \lambda & -1 + \sinh \lambda \cr
          1 + \sinh \lambda & \sinh \lambda } .
\end{equation}
From this expression, we can compute both $\lambda$ and $\theta$
in terms of the $x$ variable, as they can be written as
\begin{equation}
\sinh\lambda = x - 1 , \qquad
\sin\theta = {1 \over \sqrt{1 + (1 + x)^{2}}} .
\end{equation}

Indeed, this is an oversimplified example, but it is interesting to
note that it contains all the ingredients of the group contractions
discussed in this paper.

\section{Lorentz Group and Optical Computing}\label{comp}

Each individual is equipped with a natural computer.  He/she has ten
fingers.  With them, we can do additions and subtractions of numbers
smaller than ten. This is how our decimal system was developed.  Then
Chinese came up with the abacus which is an extension of the
ten-finger
computer.  About 150 years ago, French
artillery men invented the slide rule which converts multiplication
into addition.  In the 1940s, von Neumann observed that vacuum tubes
can perform the yes-or-no logic, and started building electronic
computers.

In building computers, it is not enough to develop computer
mathematics.
In the final stage, we have to adjust those mathematical tools
to the language spoken by devices.  As we noted in Sec.~\ref{intro},
the
Lorentz group is the standard language for classical and quantum
optics.
The Lorentz group is also the natural language for light beams and for
the materials through which the beams propagate.  Thus, if we intend
to build optical computers, we have to translate all mathematical
algorithms into the language of the Lorentz group.  In fact, it has
been
shown that some optical systems have a slide-rule-like
property~\cite{hkn97}.

In this paper, we noted first that a camera-like single-lens system
can
perform the algebra of Wigner's little groups and their contractions.
While discussing group contractions, we observed the difference
between
the rotation matrix
\begin{equation}
\pmatrix{\cos(\phi/2) & -\sin(\phi/2) \cr \sin(\phi/2) & \cos(\phi/2)}
\end{equation}
of Eq.(\ref{m200}), and the boost matrix
\begin{equation}
\pmatrix{\cosh(\lambda/2) & \sinh(\lambda/2) \cr
\sinh(\lambda/2) & \cosh(\lambda/2)}
\end{equation}
of Eq.(\ref{m208}).
These matrices operate in two different spaces, namely the rotation
matrix on a circle and the boost matrix on a hyperbola.  Since we
now have a procedure which makes an analytic continuation from one
to the other, we can perform computations in the hyperbolic world and
carry it to the circular world.

Indeed, the circle versus hyperbola is a very old problem known
as the conic sections.  It is a geometrical as well as a topological
problem, but these issues are beyond the scope of this paper.  It
is interesting to see that the single-lens system can tell us a
story about this fundamental problem.


\begin{thebibliography}{99}
\bibitem{like}
We use the word ``like'' instead of the group theoretical phrase
``locally isomorphic to," which means that the generators of two
different groups satisfy the same set of commutation relations.
The best known example is that the generators of the $SU(2)$ group
satisfy the same set of commutation relations as the three-dimensional
rotation group.
\end{thebibliography}
\end{document}